\documentclass{moriond}

\begin{document}
\vspace*{4cm}
\title{Exploring the hadronic landscape, a novel search in multijet events at the ATLAS experiment}

\author{Anthony Badea on behalf of the ATLAS Collaboration}
% \email{anthony.badea@cern.ch}
% \address{University of Chicago, Chicago, Illinois, 60637}
\address{Harvard University, Cambridge MA\\Enrico Fermi Institute, University of Chicago, Chicago IL\\Contribution to the 2024 Electroweak session of the 58th Rencontres de Moriond} % ; United States of America

\maketitle
\abstracts{
The exceptionally accurate Standard Model (SM) theory of fundamental interactions is known to be incomplete. Many new theories extend the SM, trying to solve some of the most compelling puzzles of nature. Since the start of LHC experiments, a wide range of the accessible phase space has been explored, setting robust limits on new physics. Yet, many alternative models offering less constrained final states are to be evaluated. A new search for Beyond Standard Model (BSM) physics at the ATLAS experiment in an all-hadronic final state with many jets and minimal missing energy is presented. These proceedings follow a presentation at Moriond Electroweak 2024~\cite{MoriondEW24Badea} and published results~\cite{2401.16333}.}

%  \href{https://arxiv.org/abs/2401.16333}{\textsc{2401.16333}}
%  \href{https://indico.in2p3.fr/event/32664/timetable/?view=standard_numbered#133-exploring-the-hadronic-lan}{here}

\section{Introduction}

% would like to say two points
% 1. generic motivation for all hadronic many jet final state without MET. more broadly strongly produced (via alpha_s) new physics from this being a pp collider
% 2. model specific motivation to explore r-parity violating case where run 1 limits hold for atlas
% 3. say this is with ATLAS experiment and provide citation

A new search for Beyond Standard Model (BSM) physics at the ATLAS experiment in an all-hadronic final state with many jets is presented. This signature is well motivated for a number of reasons. First, strong force production cross sections at the LHC are large because the colliding particles carry color charge. Second, the LHC collaborations have placed robust limits on signatures with missing energy and the multijet signature is under-explored due to significant challenges with the QCD background. Third, many alternative models offering less constrained final states give rise to such signatures, such as R-Parity Violating Supersymmetry (RPV SUSY), and still provide ingredients to solve outstanding questions such as the matter-antimatter asymmetry, hierarchy problem, and nature of dark matter. The search is performed using 140 fb$^{-1}$ of proton-proton ($pp$) collision data collected at $\sqrt{s}=13$ TeV by the ATLAS detector during the 2nd run of the LHC. The ATLAS experiment~\cite{PERF-2007-01} is a multipurpose particle detector with a forward--backward symmetric cylindrical geometry and a near \(4\pi\) coverage in solid angle. It consists of an inner tracking detector surrounded by a thin superconducting solenoid, electromagnetic and hadron calorimeters, and a muon spectrometer with three large superconducting air-core toroidal magnets.

\section{Analysis}

The all-hadronic state faces major challenges because the QCD interactions have the highest cross-sections at the LHC, and are remarkably complex to simulate. Two analysis strategies were developed to deal with this difficult background, a more canonical cut-and-count analysis approach and a novel search for resonances using Transformers. These methods were used to search for resonant pair production of massive particles decaying into 3 (or 5) SM quarks each. SUSY gluinos decaying via RPV couplings were considered as benchmark signal models. The results show a strong improvement in sensitivity with respect to previous ATLAS searches, beyond simple luminosity scaling.

\begin{figure}[t]
    \centering
    \includegraphics[width=0.48\textwidth,trim={0cm 0cm 0 0},clip]{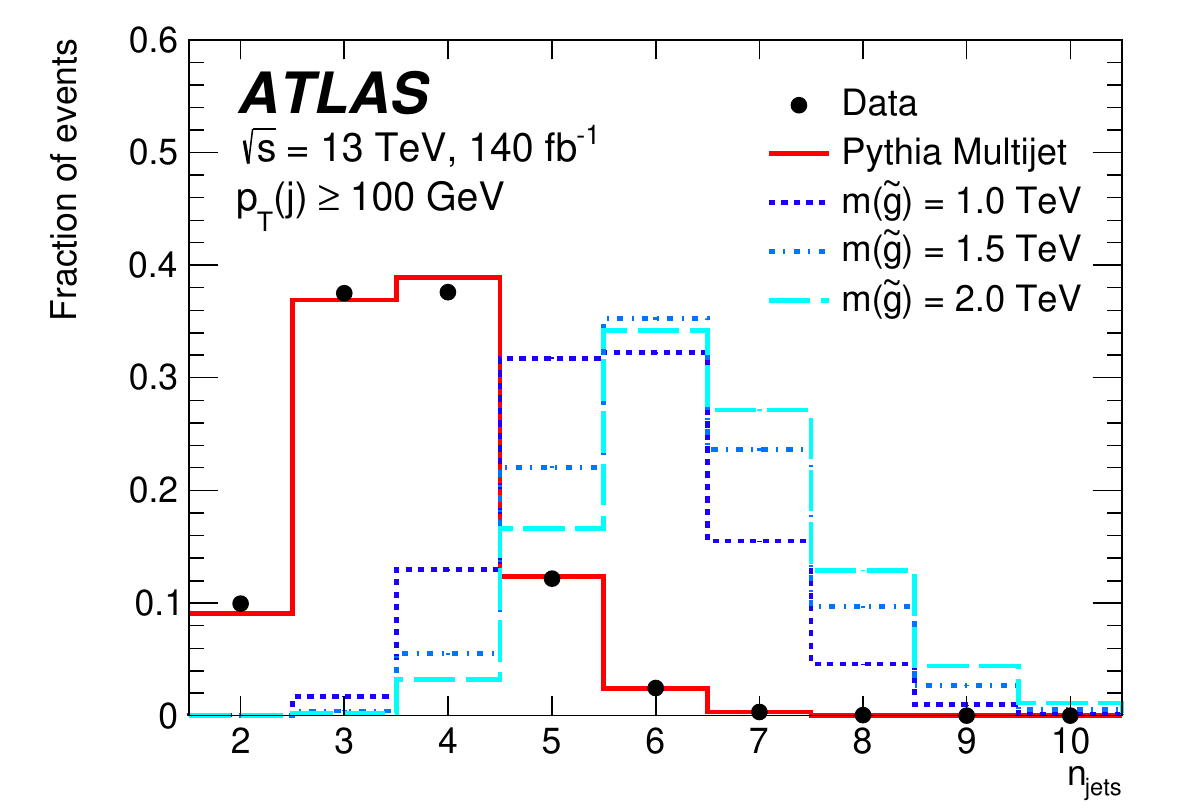}
    \includegraphics[width=0.48\textwidth,trim={0cm 0cm 0 0},clip]{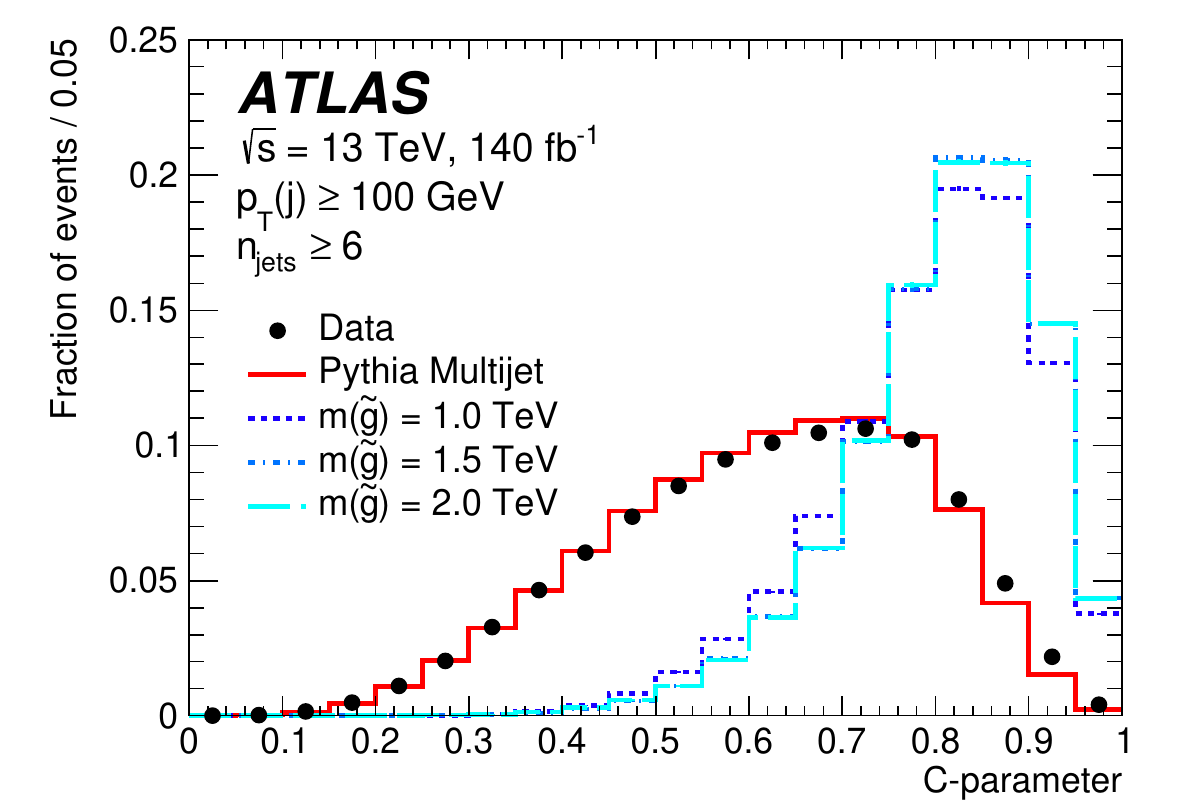}
    \caption{
      Comparison between distributions, normalised to unity,  of the observed data, the QCD multijet background and signal models.
      Left: $n_{\textrm{jets}}$ spectrum with $p_{\mathrm{T}} \geq 100$ GeV.
      Right: distribution of the $C$ variable for events with at least six jets with $p_{\mathrm{T}} \geq 100\,$ GeV.
    }
    \label{fig:var-shapes}
  \end{figure}

The targeted signal is pair produced high mass particles that decay to a large number of quarks, leading to many jets. To produce heavy particles, the colliding partons must originate from the energetic tails of the parton distribution functions (PDF). This means that there is a lot of energy available to produce high $p_{\mathrm{T}}$ jets. In comparison, pure QCD production leads to softer jets because the PDFs are steeply falling. Furthermore, since the collisions are from the tails of the PDFs the heavy particles are typically produced close to rest, leading to isotropically distributed decay products. These two features are quantified in the n$_{\mathrm{jets}}$ and $C$-parameter distributions shown in Fig. \ref{fig:var-shapes}. The $C$-parameter is an event shape variable derived from the eigenvalues of the linearized sphericity tensor. Both analysis methods place cuts on the variables.

% Since the resonances are heavy, there is  leads to a larger number of high-$p_{\mathrm{T}}$ jets and more isotropic energy spread per event in comparison to the QCD background. 

% because there is a lot of energy available from the high mass resonance and the final state is quark enriched. 

\paragraph{Jet Counting Method Summary}

% would like to say two points:
% 1. here is how the jet counting method (strategy + background estimate) is done
% 2. here is how the mass resonance method is done

Seven signal regions requiring at least 7 jets with $p_{\mathrm{T}} >$ 180 GeV and $C \geq 0.85$ are used. Two of the regions require at least 2 $b$-tagged jets to enhance sensitivity to heavy flavour decays. The background is estimated from events containing low jet multiplicities and low momenta extrapolated to higher jet-momenta and multiplicities using MC scale factors. To compensate the limitations of the simulated multi-jet background, data are used to normalize the simulation in several control regions.

\paragraph{Mass Resonance Method Summary} The objective of the mass resonance method is to observe a resonance in the reconstructed candidate gluino mass spectrum. In contrast to the jet counting method, which could have an excess from a variety of high energy contributions, this search would be an unambiguous sign of new physics from resonant production at the probed energy scale. For the direct gluino decay model, the combinatorial issue of correctly identifying which jets should be matched to each gluino candidate is a significant problem and the main objective of the design of the method. A dedicated neural network (NN) is developed using the attention mechanism from the transformer architecture to correctly group together the jets from each individual gluino candidate decay. Machine-learning techniques were applied previously to the combinatorial assignment problem focusing on the reconstruction of standard model processes and with less focus on beyond Standard Model scenarios given the additional unknown of the exotic particle masses. The invariant mass $m_{\tilde{g},i}$ of the two gluino candidates is built from the jets that are selected by the NN and the average of the two masses $m_{\mathrm{avg}} = \frac{1}{2} \left(m_{\tilde{g},1} + m_{\tilde{g},2}\right)$ is used as the key discriminating variable. The method searches for a localised excess on the $m_{\mathrm{avg}}$ spectrum where the background is estimated through a functional fit to a smooth decreasing spectrum.

Events used by the mass resonance method require at least six jets with $p_{\mathrm{T}}$ above 100 GeV, and $C \geq 0.9$. A second selection is defined requiring in addition at least one $b$-tagged jet, which is used to improve the sensitivity to heavy flavour decays. Further to the previously introduced selections, which are used for the model-dependent interpretation, a set of model-independent SRs are defined using single bins in the invariant mass distribution with a width of 300 GeV, and assume no signal contribution outside of the SR.

\begin{figure}[t]
  \centering
  \includegraphics[width=0.48\textwidth]{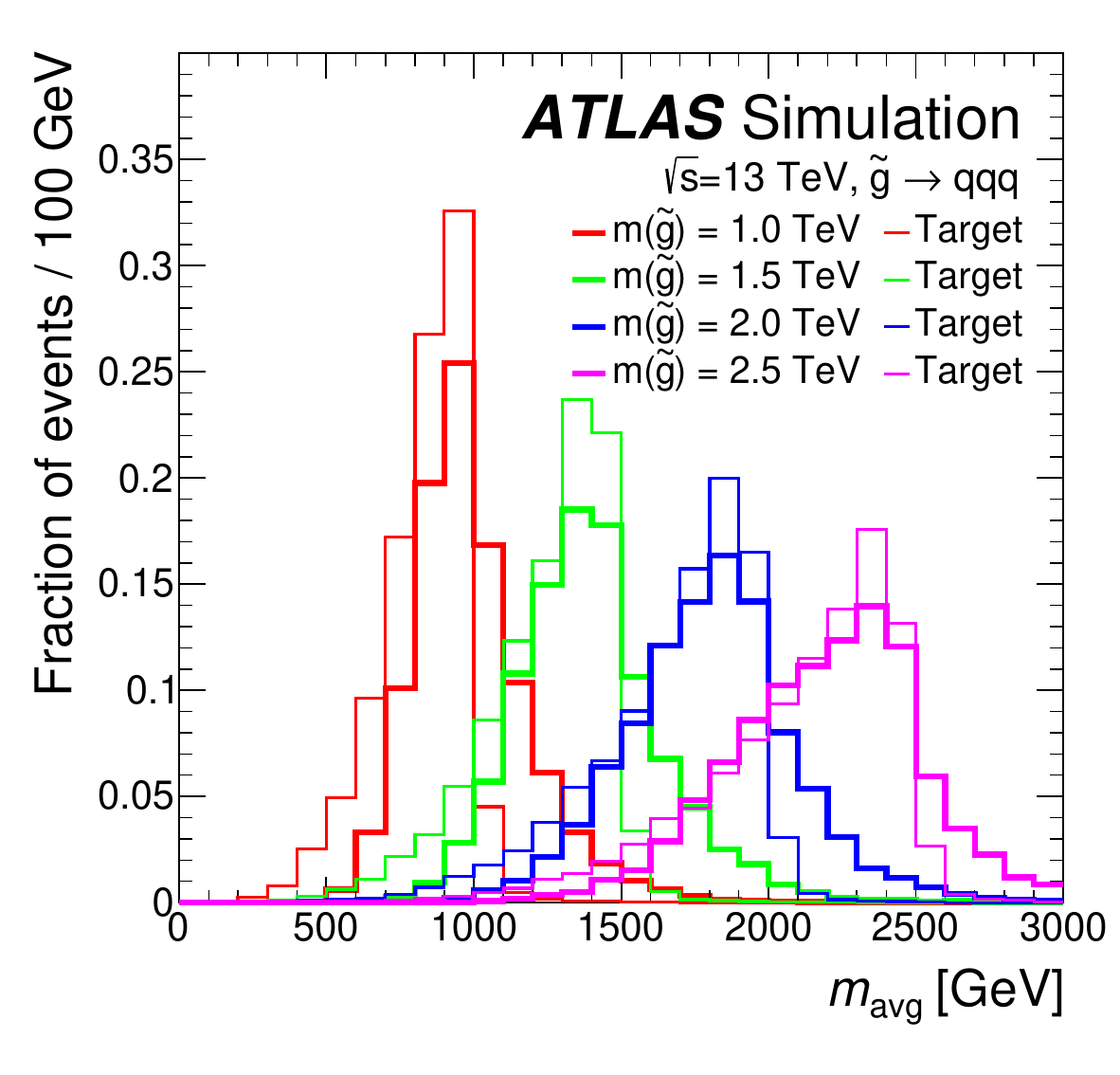}
  \includegraphics[width=0.48\textwidth]{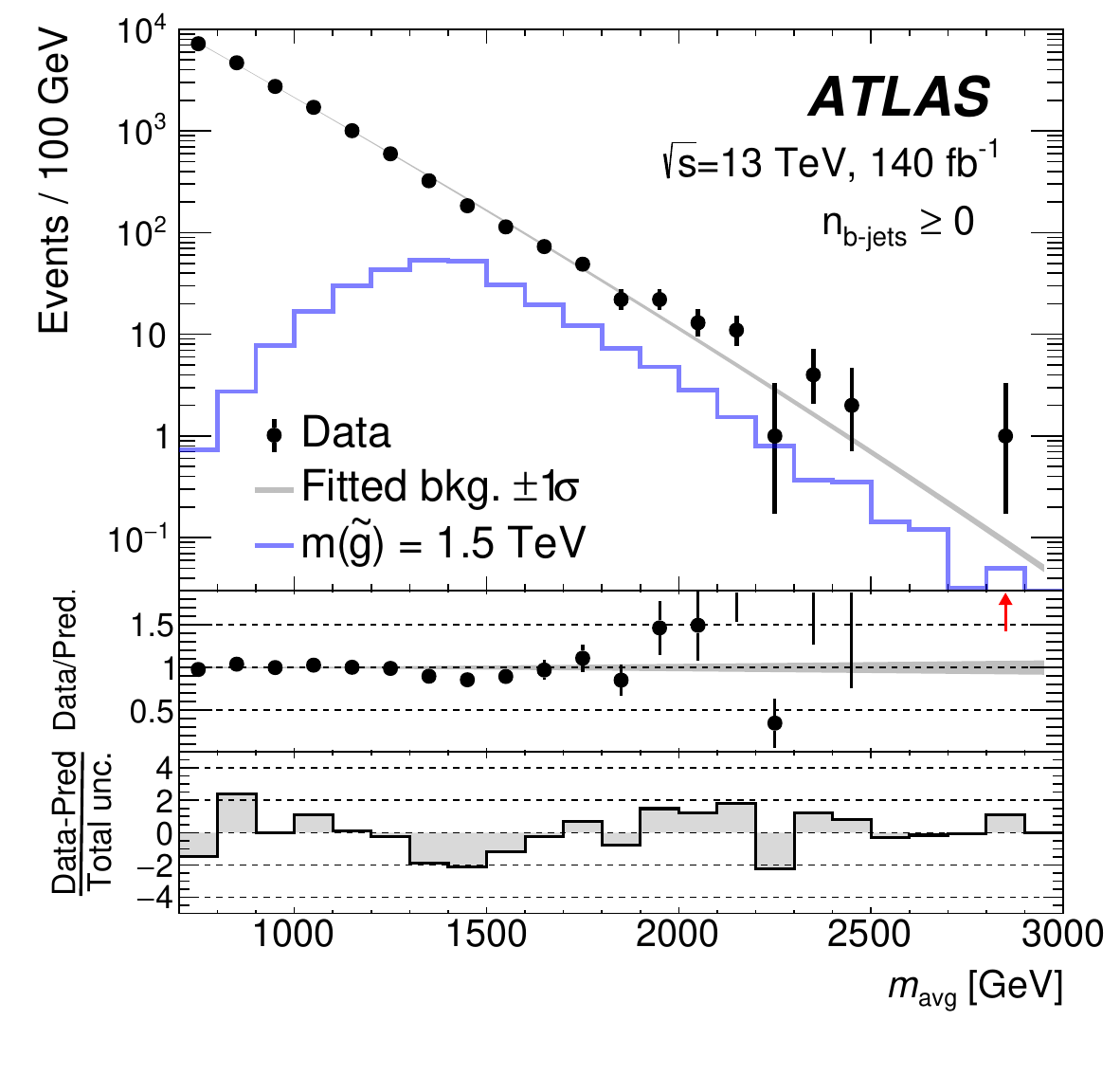}
  \caption[]{ Left: normalised average mass spectrum comparing the shapes of the reconstructed (solid) and target (light) distributions for different masses. The reconstructed distribution is produced using the NN assignments, whereas the target distribution is built assigning jets to gluinos based on their MC generated labels. Right: background-only fit to the reconstructed average mass spectrum of the candidate gluinos, in the nominal and regions of the mass resonance method. The grey bands include both the statistical and systematic uncertainties. The red arrow denotes points that lie above the range of the ratio plot.
  
  % Background-only fits to the reconstructed average mass spectrum of the candidate gluinos in (left) the nominal region of the mass resonance method.
% The grey bands include both the statistical and systematic uncertainties.
% The red arrow denotes points that lie above the range of the ratio plot.
% Observed and expected 95\% CL upper limits on the signal cross-section times branching ratio ($\sigma \;\times$ BR) as a function of the gluino mass for the gluino direct decay model with \UDS decays.
% The expected limits for the jet counting and mass resonance methods are shown in red and blue, respectively. The best expected limit per mass point between the methods is chosen (dashed black) and corresponding observed limit reported (solid black).
% The green and yellow bands around the expected limit correspond to the $\pm 1 \sigma$ and $\pm 2 \sigma$ variations including both the systematic and statistical uncertainties, respectively.
% The theoretical prediction is also shown, with the uncertainties in the prediction shown as a coloured band.  
  }
  \label{fig:ml}
\end{figure}

The performance of the NN is illustrated on the left in Fig.~\ref{fig:ml}. The reconstructed mass matches the target with a small loss in resolution as expected. Target signals show a low-mass tail, especially at higher masses, which originates from the restriction that exactly three jets are matched. This requirement misses additional final state radiation jets which are significant for the highest masses. The inclusive (no $b$-tagging requirement) SR is shown on the right in Fig.~\ref{fig:ml}. As mentioned, the method is used to set model-independent limits but since the NN is trained on signal MC, the method is inherently model-dependent.

\section{Results}

\begin{figure}[t]
  \centering
  \includegraphics[width=0.65\textwidth]{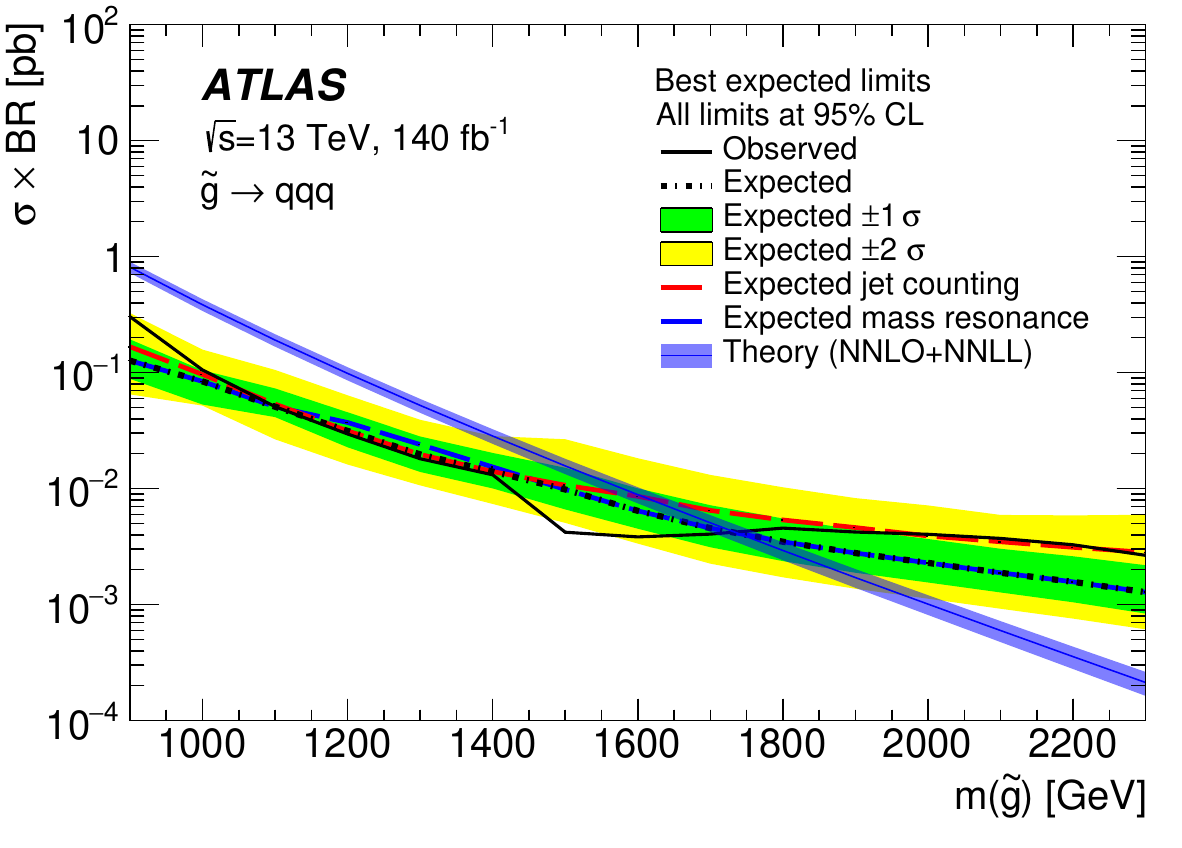}
  \caption[]{
    Observed and expected 95\% CL upper limits on the signal cross-section times branching ratio ($\sigma \;\times$ BR) as a function of the gluino mass for the gluino direct decay model with \textit{UDS} decays.
    The expected limits for the jet counting and mass resonance methods are shown in red and blue, respectively. The best expected limit per mass point between the methods is chosen (dashed black) and corresponding observed limit reported (solid black).
    The green and yellow bands around the expected limit correspond to the $\pm 1 \sigma$ and $\pm 2 \sigma$ variations including both the systematic and statistical uncertainties, respectively.
	The theoretical prediction is also shown, with the uncertainties in the prediction shown as a coloured band.
  }
  \label{fig:limit}
\end{figure}

No significant excess of data over the expected event yields is observed in any signal region. Limits are set on the production of gluinos in the gluino direct decay and cascade decay models in $U\bar{D}\bar{D}$ scenarios of RPV SUSY. Fig. \ref{fig:limit} shows the exclusion curve for the inclusive direct decay model. The mass resonance method (blue) is sensitive to roughly $3$ times lower $\sigma\times\mathrm{BR}$ than the jet counting method (red). From naive scaling of $S/\sqrt{B}$ this means that the jet counting method would need roughly 9 times more data to achieve similar sensitivity. The final exclusion uses the method with the best expected performance per mass point to be less sensitive to data fluctuations. The gain in performance is smaller for the $b$-tagged SR, where there is less new information for the NN to utilize. The cascade decay limits are entirely computed using the jet counting method. In the gluino direct decay model, gluinos with masses up to 1800 GeV are excluded at 95\% CL. In the gluino cascade decay model, gluinos with masses as high as 2340 GeV are excluded for a neutralino with 1250 GeV mass. Model-independent limits are also set on the visible cross-section times branching ratio in five overlapping signal regions. 

% would like to say two points:
% 1. world leading constraints on this particular model
% 2. look at improvement from using mass resonance method

\section{Conclusion}
A new search for BSM physics in an all-hadronic final state with many jets is presented. These results improve upon the previously existing LHC limits owing to the larger luminosity, the introduction of event shape variables to suppress background, and the development of machine-learning techniques to assign jets to gluinos and reconstruct their mass. The techniques developed for this analysis pave the way for future searches and measurements in high multiplicity final states and all-hadronic scenarios.

% A search for R-parity-violating SUSY signals in events with multiple jets is performed with 140 fb $^{-1}$ of proton–proton collision data at $\sqrt{s} = 13$ TeV collected by the ATLAS detector at the LHC. 

% Two methods are used, a jet counting method searching for excess events in single-bin signal regions defined at high jet multiplicity and high $C$, and a mass resonance approach, which searches for a localised excess in the reconstructed gluino mass spectrum. A novel machine-learning approach is employed to address the combinatorial assignment problem and successfully reconstruct the gluino mass. 

\section*{Acknowledgments}

Thank you to Stefano Franchellucci for the helpful comments. This work has been supported by the Department of Energy, Office of Science, under Grant No. DE-SC0007881, Harvard Graduate Prize Fellowship, and Eric and Wendy Schmidt AI in Science Fellowship.

% \section*{Appendix}

\section*{References}
\bibliography{Badea}

\begin{thebibliography}{1}

\bibitem{MoriondEW24Badea}
Exploring the hadronic landscapes, a novel search in multi-jet events in atlas.
\newblock
  \url{https://indico.in2p3.fr/event/32664/timetable/?view=standard_numbered#133-exploring-the-hadronic-lan}.

\bibitem{2401.16333}
ATLAS collaboration.
\newblock A search for $\mathrm{R}$-parity-violating supersymmetry in final
  states containing many jets in $pp$ collisions at $\sqrt{s} =
  13\;\mathrm{TeV}$ with the $\mathrm{ATLAS}$ detector.
\newblock {\em Journal of High Energy Physics}, 2024(5):3, 2024.

\bibitem{PERF-2007-01}
{ATLAS Collaboration}.
\newblock {The ATLAS Experiment at the CERN Large Hadron Collider}.
\newblock {\em JINST}, 3:S08003, 2008.

\end{thebibliography}

\end{document}